\pdfoutput=1
\documentclass{PoS}

\title{Higgs Search in $b\bar{b}$ Signatures at ATLAS and CMS}

\ShortTitle{Higgs Search in $b\bar{b}$ Signatures at ATLAS and CMS}

\author{\speaker{Yoshikazu Nagai}\\
        CPPM, Aix-Marseille Universit\'e\\
        E-mail: \email{Yoshikazu.Nagai@cern.ch}\\
        on behalf of the ATLAS and CMS Collaborations}

\abstract{
We report on searches for the Standard Model Higgs boson 
decaying to $b \bar{b}$ in $p p$ collisions at the ATLAS and CMS experiments.
These rely on signatures where the Higgs boson is produced in association with 
a vector boson or in association with a top-quark pair.
Results are presented, based on $5.0~\mathrm{fb}^{-1}$ of $pp$ collision data at 
$\sqrt{s} = 7$~TeV and up to $13.0~\mathrm{fb}^{-1}$ of data at $\sqrt{s} = 8$~TeV 
collected by the LHC accelerator. 
These are expressed in terms of 95\% confidence level upper limits on the
production cross section of the Standard Model Higgs boson times 
the branching ratio to decay to $b$-quark pair.
}

\FullConference{14th International Conference on B-Physics at Hadron Machines,\\
		April 8-12, 2013\\
		Bologna, Italy }

\usepackage{graphicx}

\begin{document}


\section{Introduction}
\vspace{-1.5mm}
In July 2012, the observation of a Higgs-like boson was announced with a mass around 
125~GeV by both the ATLAS~\cite{ATLAS} and CMS~\cite{CMS} collaborations, 
relying mainly on the bosonic decay modes ($\gamma\gamma$, $WW$, $ZZ$) of the new boson.
Recently, both the ATLAS and CMS have updated Higgs boson search results~\cite{ATLAS_RecentHiggs,CMS_RecentHiggs}.
Now that the observation of the new boson is established,
the important question to be addressed is: 
Whether is this new boson the Standard Model (SM) Higgs boson or something else?

To confirm the nature of this new boson
both ATLAS and CMS experiments started to measure the properties of this boson 
such as mass, spin, parity and couplings.
In addition, if it is the SM Higgs boson, 
it predominantly decays to $b$-quark pairs (Br($H \to b \bar{b}$)$\simeq 58\%$) 
at the mass of 125~GeV. 

In this proceeding, we present the searches for the SM Higgs boson decaying to 
$b$-quark pairs with the Higgs boson produced in association with a $W$ or a $Z$ boson 
(denoted as $VH$) or with the Higgs boson produced in association with top-quark pairs
(denoted as $ttH$).

\vspace{-3.5mm}
\section{Search for the Higgs Boson in the $VH$, $H \to b\bar{b}$}
\vspace{-1.5mm}
The gluon fusion and the vector boson fusion production process have 
the first and second highest production cross section at the LHC, 
however,  $H \to b\bar{b}$ analyses exploiting these production modes are quite
difficult due to the overwhelming QCD multi-jet background.
On the other hand, the QCD multi-jet background can be efficiently suppressed 
exploiting the $VH$ production mode, selecting events based on
the leptonic decay modes of the associated vector boson. 
The $WH$ and $ZH$ production modes have the third and fourth
highest production cross section: 
$\sigma(WH) = $~0.697~pb (0.573~pb), $\sigma(ZH) = $~0.394~pb (0.316~pb) 
for $\sqrt{s} = $8~TeV (7~TeV) at $m_H = $125~GeV, respectively~\cite{LHC_Xsection}. 

Three exclusive categories are defined based on the leptonic decay signature, 
selecting predominantly $ZH \to \nu \nu b\bar{b}$ (0-lepton), 
$WH \to \ell \nu b\bar{b}$ (1-lepton) and $ZH \to \ell \ell b\bar{b}$ 
(2-lepton)\footnote{$\ell$ denotes an electron or a muon.}.
To maximize the Higgs boson discovery potential, 
we further sub-divide the signal regions based on the vector boson $p_{\rm T}$.
The presence of a signal (i.e. an excess of data events above the background expectation) 
is tested in both ATLAS and CMS through a maximum likelihood fit to the signal 
and background normalizations, 
based on a discriminating variable and several different event categories. 
In ATLAS the discriminating variable has been chosen to be the invariant mass 
of the 2 b-jets ($m_{b\bar{b}}$), 
while CMS uses the shape of the output discriminant of a 
Boosted Decision Tree (BDT) algorithm which takes as input several kinematic variables 
including $m_{b\bar{b}}$ and $b$-tagging information.
Since final sensitivity relies heavily on the $m_{b\bar{b}}$ invariant mass resolution,
both ATLAS and CMS analysis implement the way to improve the $b$-jet energy resolution.
The analyzed datasets correspond to an integrated luminosity of 
$13.0~\mathrm{fb}^{-1}$ ($4.7~\mathrm{fb}^{-1}$) in $pp$ collisions 
at $\sqrt{s} = 8$~TeV ($\sqrt{s} = 7$~TeV) in ATLAS and an integrated luminosity of 
$12.1~\mathrm{fb}^{-1}$ ($5.0~\mathrm{fb}^{-1}$) in CMS.

In the following sub-sections these analyses are presented in more detail. 
Full details are given in the conference notes 
of the ATLAS~\cite{ATLAS_VH} and CMS analyses~\cite{CMS_VH}.

\subsection{Event Selection}
The baseline event selection is summarized in Table~1 
for both the ATLAS and CMS analyses. 
The 1-lepton and 2-lepton channels require to pass a single high-$p_{\rm T}$ lepton 
trigger or di-lepton trigger. 
In the 0-lepton channel the online event selection is based 
in ATLAS on a pure $E_{\rm T}^{\rm miss}$ trigger, 
while in the CMS analysis the $E_{\rm T}^{\rm miss}$ signature is completed by the 
presence of jets in the event with or without an additional requirement on the 
vectorial sum of transverse momenta of the physics objects in the event. 
The requirements on the leptons vary across channels as outlined in Table~1. 
Jets are reconstructed using an anti-$k_{\rm T}$ clustering algorithm~\cite{antikt} 
with radius parameter $R = $0.4 for the ATLAS analysis 
and reconstructed from particle-flow objects 
using an anti-$k_{\rm T}$ clustering algorithm with $R = $0.5 
for the CMS analysis.
To identify jets originating from $b$-quarks, 
the ATLAS analysis uses the MV1 algorithm~\cite{MV1} which combines several informations
from different algorithms using an artificial neural network algorithm with an efficiency of 70\% for $b$-jets 
and with an efficiency of 0.7\% for $light$-jets, 
while the CMS analysis uses the Combined Secondary Vertex (CSV) $b$-tagging algorithm~\cite{CSV}
with an approximate efficiency of 50\% (72\%) for $b$-jets and with an efficiency of 0.15\% (3\%) for $light$-jets
for tight (loose) $b$-tag.
In addition to the baseline event selection, further topological cuts are applied to maximize 
the analysis sensitivity, such as angular cuts between the two leading jets.

\begin{table}[t]
  \label{table:1}
  \caption{
    The baseline event selection.
    The complete event selection can be found in reference~\cite{ATLAS_VH} 
    for the ATLAS analysis and in reference~\cite{CMS_VH} for the CMS analysis. 
  }
  \begin{center}
    \begin{tabular}{c|c|c|c}\hline \hline
      & 0-lepton & 1-lepton & 2-lepton \\ \hline \hline    
      Lepton & 0 loose & 1 tight + 0 loose & 1 medium + 1 loose \\ \hline
      Jets & \multicolumn{3}{c}{$p_{\rm T}^1 > 45$ GeV, $p_{\rm T}^{2,3} > 20$ GeV, $|\eta| < 2.5$ (ATLAS)} \\ 
           & \multicolumn{3}{c}{$p_{\rm T}^1 > 60/30/20$ GeV, $p_{\rm T}^{2} > 30/30/20$ GeV (0/1/2 leptons), $|\eta| < 2.5$ (CMS)} \\ \hline
      Number of jets & $=$ 2 or 3 jets (ATLAS) & \multicolumn{2}{c}{$=$ 2 jets (ATLAS)}\\ 
                     & \multicolumn{3}{c}{$\geq$ 2 jets (CMS)}  \\ \hline
      $b$-tag & \multicolumn{3}{c}{2 $b$-jets (ATLAS)} \\ 
      & \multicolumn{3}{c}{btag1: medium-loose for all channel (CMS)}   \\ 
      & \multicolumn{3}{c}{btag2: loose/medium-loose/loose (0/1/2 leptons) (CMS)}   \\ \hline 
      $E_{\rm T}^{\rm miss}$ & $> 120$ GeV(ATLAS) & $> 25$ GeV (ATLAS)& $< 50$ GeV (ATLAS)\\
                             & $> 130$ GeV (CMS) & $> 45$ GeV (CMS) & -- (CMS)\\ 
\hline \hline
    \end{tabular}
  \end{center}
\end{table}

After selection the events are sub-divided into several signal regions based on the 
$p_{\rm T}$ of the $V$ boson and number of jets: as a result the ATLAS
analysis has 16 signal regions across the three channels, while the CMS analysis has 10 signal regions.
In the high $p_{\rm T}(V)$ regions the two $b$-jets from the Higgs boson decay are expected to be collimated, 
while this is not true for example for the top pair background.
Therefore, topological cut values are optimized depending on the $p_{\rm T}(V)$ region considered.


%

\subsection{Background Estimation and Analysis Optimizations}
The dominant backgrounds after event selection are $W$+jets, $Z$+jets and top production.
The normalization for the most background distributions are estimated
using control regions in data with the shape taken from simulation.
The two exceptions are the QCD multi-jet background, 
which is estimated entirely from data, and the diboson ($WW$, $WZ$, $ZZ$) background 
for which both the normalization and the shape are taken from simulation.
Typical control regions rely on selecting events outside the Z mass window 
(2-lepton channel) or requiring one additional jet to be present
in the event (1-lepton channel) to constrain top production, 
or on selecting events outside the $m_{b\bar{b}}$ signal window to constrain
$V$+$b$/$c$/$l$-jet production, where the $b$/$c$/$l$ flavor composition is 
constrained by exploiting events with zero, one or two $b$-tagged jets.

To further improve the mass resolution, both the ATLAS and CMS analyses 
develop $b$-jet energy correction method with different approach.
In the ATLAS analysis, the mass resolution is improved 
by adding the energy from muons within the jet to the total jet energy and 
accounting for the missing contributions from neutrinos escaping detection.
A $p_{\rm T}$-dependent correction is also applied to account for biases 
in the response caused by resolution effects of the jets from the Higgs boson decay. 
The improvement on the mass resolution is approximately 6\%.
In the CMS analysis, the mass resolution is improved by applying regression techniques 
similar to those used at the CDF experiment~\cite{CDF_bjetECorr}.
For this purpose, a specialized BDT algorithm is trained on simulated $H \to b\bar{b}$
signal events with inputs that include detailed information about the jet structure 
and that help differentiate jets from $b$-quarks from $light$-flavor jets. 
The improvement on the mass resolution is approximately 15\%.

\subsection{Diboson Observation} 
Diboson production with a $Z$ boson decaying to $b\bar{b}$ and such $Z$ boson produced 
in association with either a $W$ or another $Z$ boson presents exactly the final state 
as the $VH$ signal, 
but with a cross section around five times larger than $VH$ production.
Despite the fact that in diboson production the $p_{\rm T}(V)$ spectrum is significantly 
softer than for $VH$ production, the search for the well-known SM diboson signal allows 
to validate the signal extraction methods used in the $VH$ analyses.

The statistical analysis of the data employs 
a binned likelihood function $\mathcal{L}(\mu, \theta)$
constructed as the product of Poisson probability terms 
for each category of events and the product over all data events of 
the sum of the signal and background probability density functions (PDFs) 
for the discriminating variable, where $\theta$ is nuisance parameters to 
describe systematic uncertainties.
A signal strength parameter, $\mu$, 
multiplies the expected diboson production cross section.
Exclusion limits are set based on the {\it CLs} method~\cite{CLs_1,CLs_2} and
a value of $\mu$ is regarded as excluded when {\it CLs} takes 
on the corresponding value.

Figure~1 shows the distribution in data 
after subtracting all background evaluated after the profile likelihood fit 
except the $WZ$ and $ZZ$ diboson contributions for the ATLAS and CMS analyses. 
Clear excesses in the data compared to the background are observed
around the expected mass position of the vector boson signal.
After combining all the channels, 
ATLAS observes $\mu_D = {\rm 1.09} \pm 0.20 ({\rm stat.}) \pm 0.22 ({\rm syst.})$,
which corresponds to the significance of 4.0\,$\sigma$ and agrees with 
the SM prediction value of $\mu_D = 1$.  
CMS also observes clear excess with a rate approximately as expected from the standard model.

\begin{figure}[htbp]
  \label{fig1}
  \begin{tabular}{c}
    \begin{minipage}{0.55\hsize}
      \begin{center}
        \vspace{0.5cm}
        \includegraphics[width=0.8\textwidth,clip]{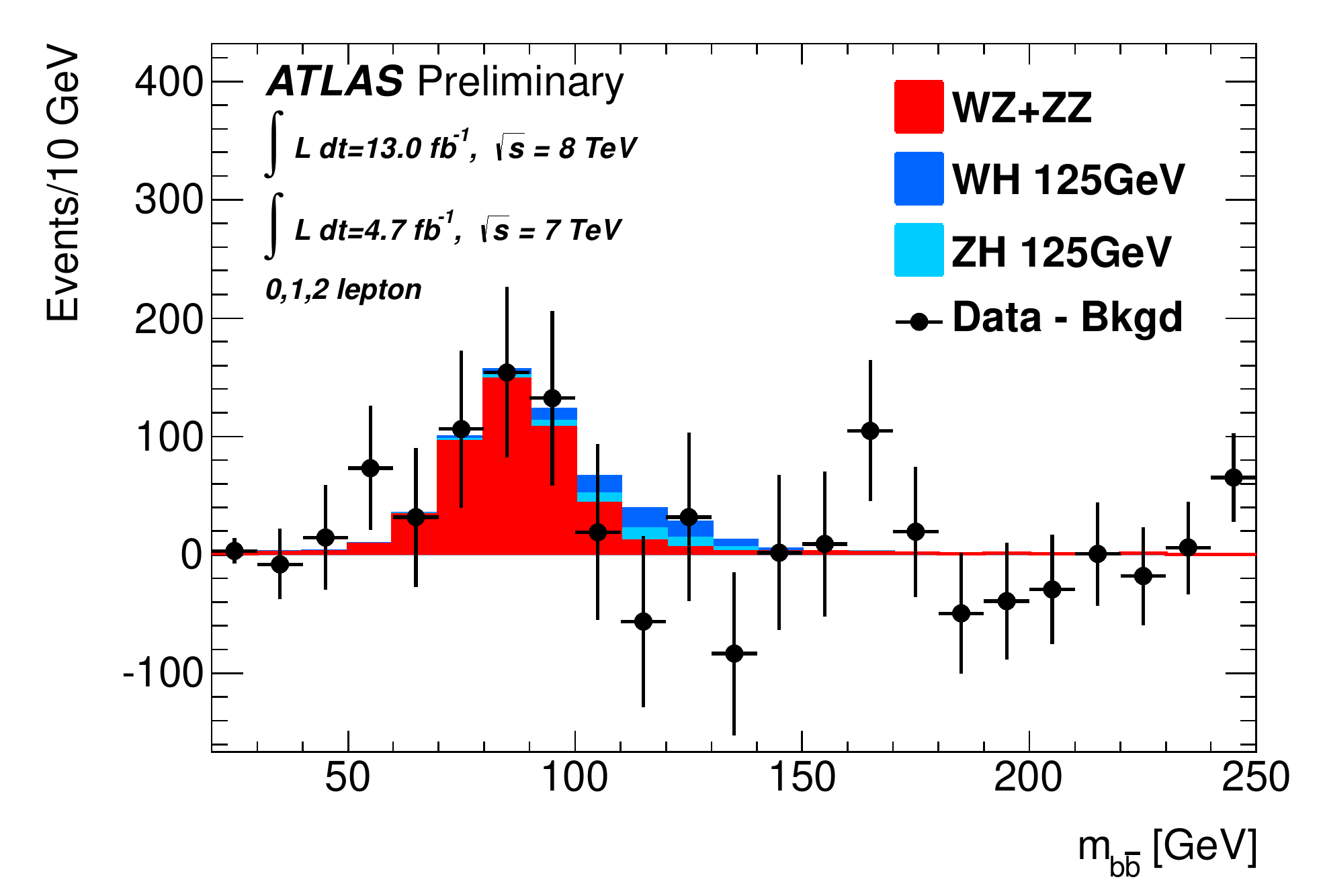}
      \end{center}
    \end{minipage}

    \begin{minipage}{0.45\hsize}
      \begin{center}
        \includegraphics[width=0.75\textwidth,clip]{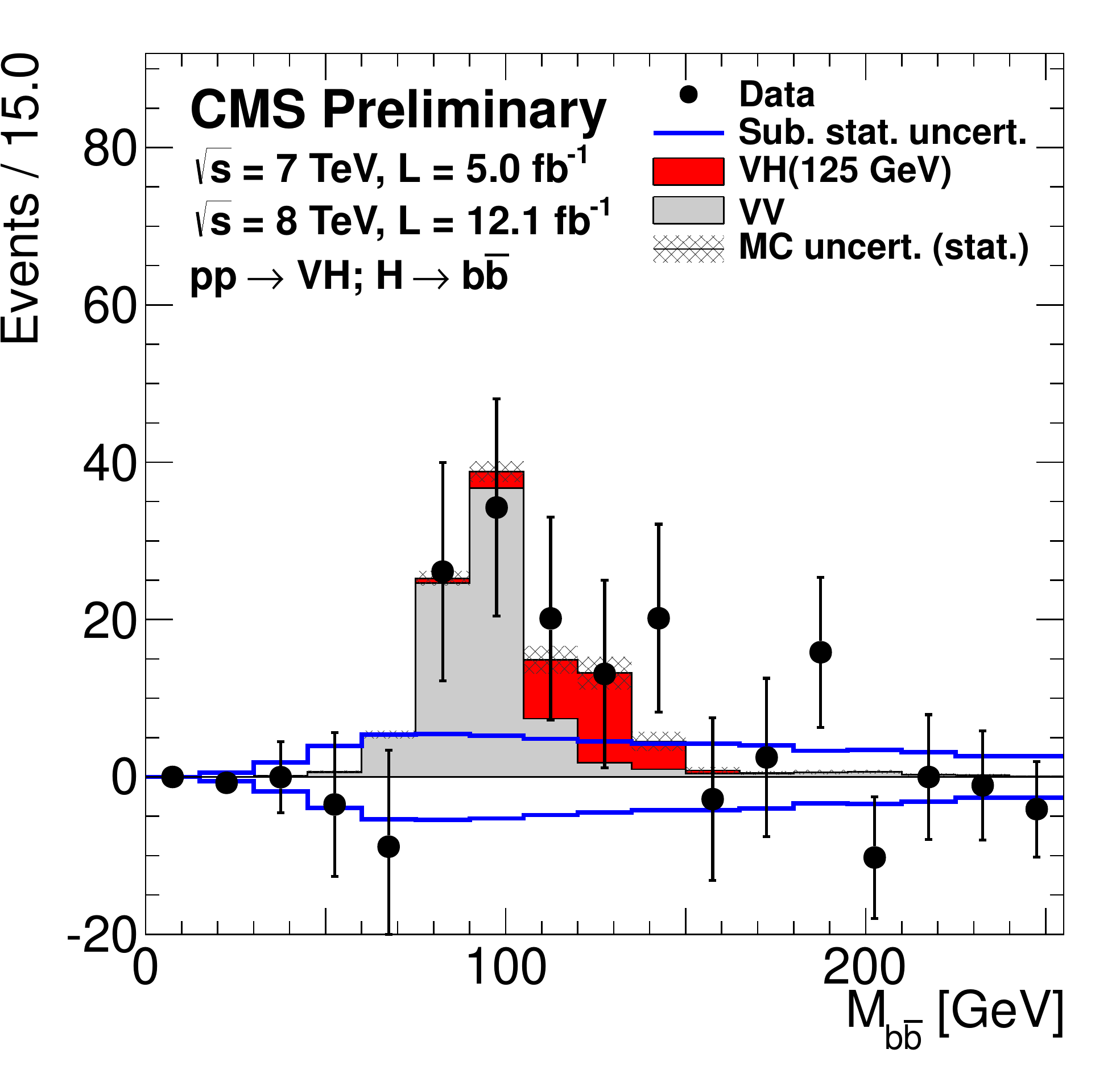}
      \end{center}
    \end{minipage}
    
  \end{tabular}  

  \caption{$m_{b\bar{b}}$ distribution in data after subtraction of all background
    except the $WZ$ and $ZZ$ diboson contributions 
    including expected SM Higgs boson signal.
    (Left) ATLAS analysis~\cite{ATLAS_VH}. (Right) CMS analysis~\cite{CMS_VH}.
  }
\end{figure}

\subsection{Results} 
In order to test the presence of the Higgs boson signal,
the test statistic is constructed similarly as diboson analysis discussed 
in the previous sub-section.
The ATLAS analysis uses the shape of the $m_{b\bar{b}}$ distribution for the fit, 
while the CMS analysis uses the shape of the output of the BDT algorithms trained separately 
for each channel and for each Higgs boson mass hypothesis.
Dominant systematic uncertainties are $b$-tagging calibration uncertainties, 
jet energy scale, missing transverse energy scale, MC statistics and modeling.
Figure~2 shows examples of $m_{b\bar{b}}$ distributions from the ATLAS analysis
and Fig.~3 shows examples of BDT algorithm output distributions from the CMS analysis. 

\begin{figure}[htbp]
  \label{fig2}
\begin{center}
         \includegraphics[width=0.43\textwidth,clip]{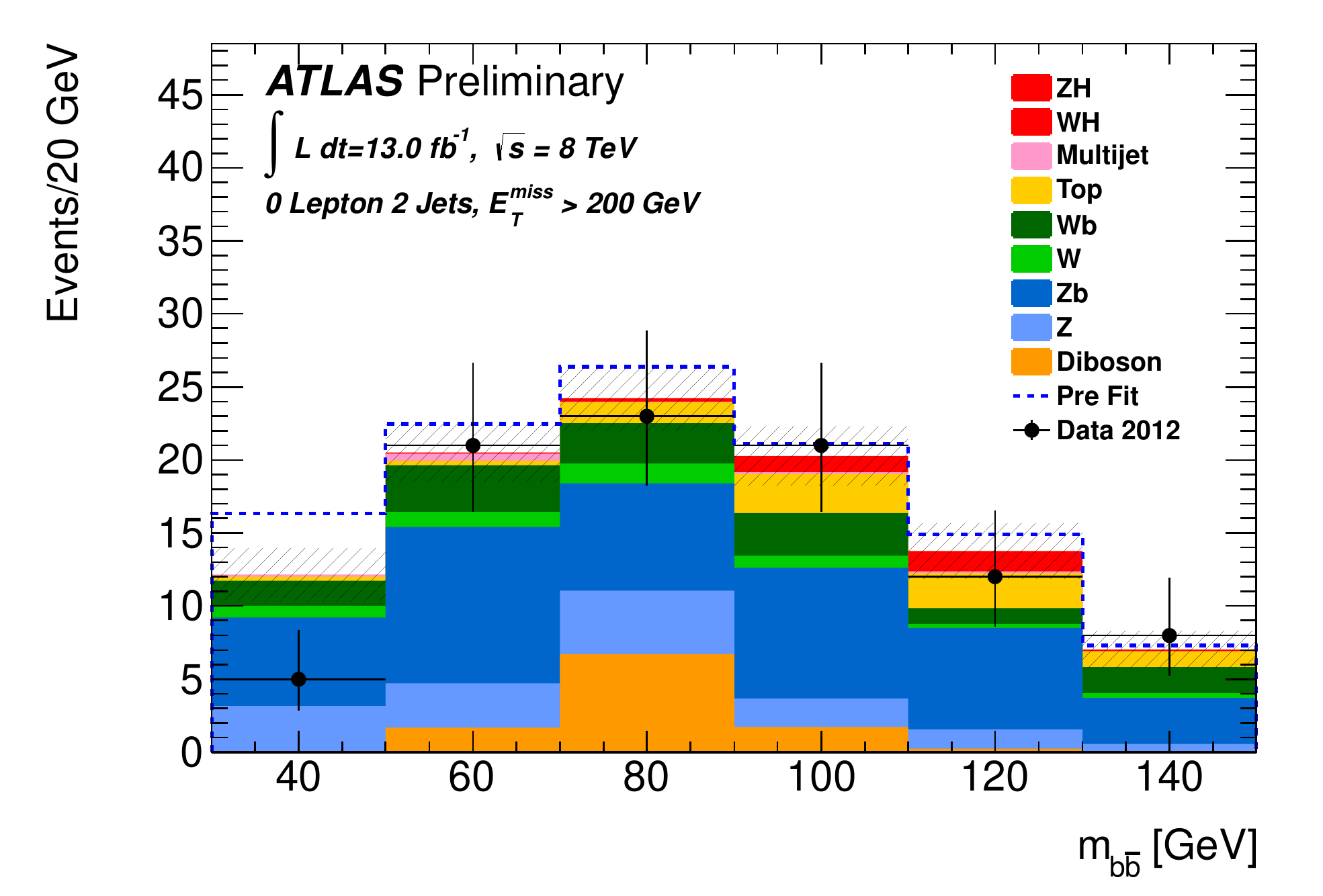}
         \includegraphics[width=0.43\textwidth,clip]{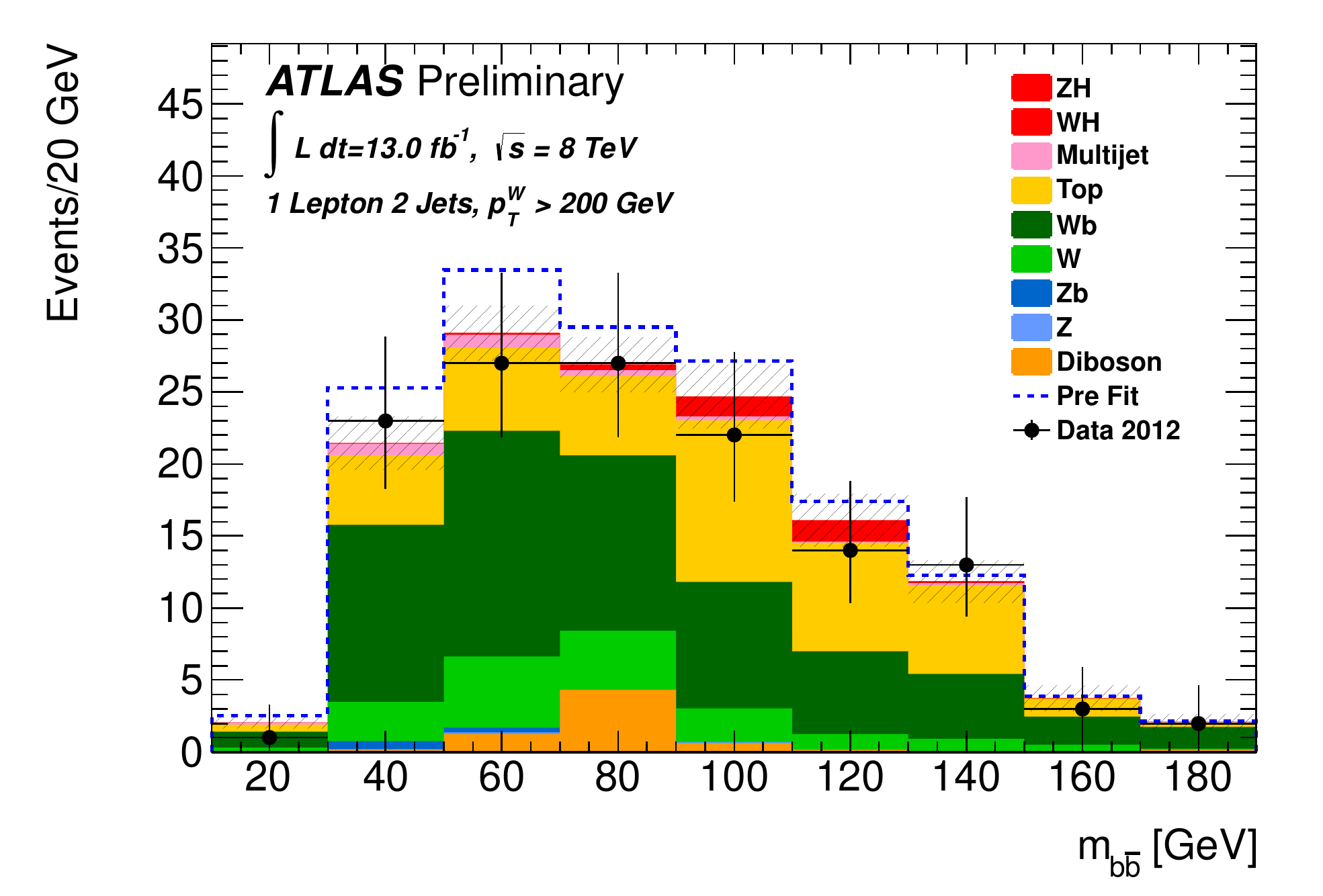}
  \caption{Examples of $m_{b\bar{b}}$ distributions 
    for the ATLAS $VH$ analysis~\cite{ATLAS_VH}.
    The signal in the plots is for $m_H = $125~GeV.
    The background expectation is shown after the profile likelihood fit (solid) 
    and compared to the predictions from the pre-fit Monte Carlo simulation (dashed).
    (Left) 0-lepton channel, 2-jet, $E_{\rm T}^{\rm miss} > $200~GeV. 
    (Right) 1-lepton channel, 2-jet, $p_{\rm T}^W > $200~GeV. 
  }
\end{center}
\end{figure}

\begin{figure}[htbp]
  \label{fig3}
  \begin{center}

  \includegraphics[width=0.34\textwidth,clip]{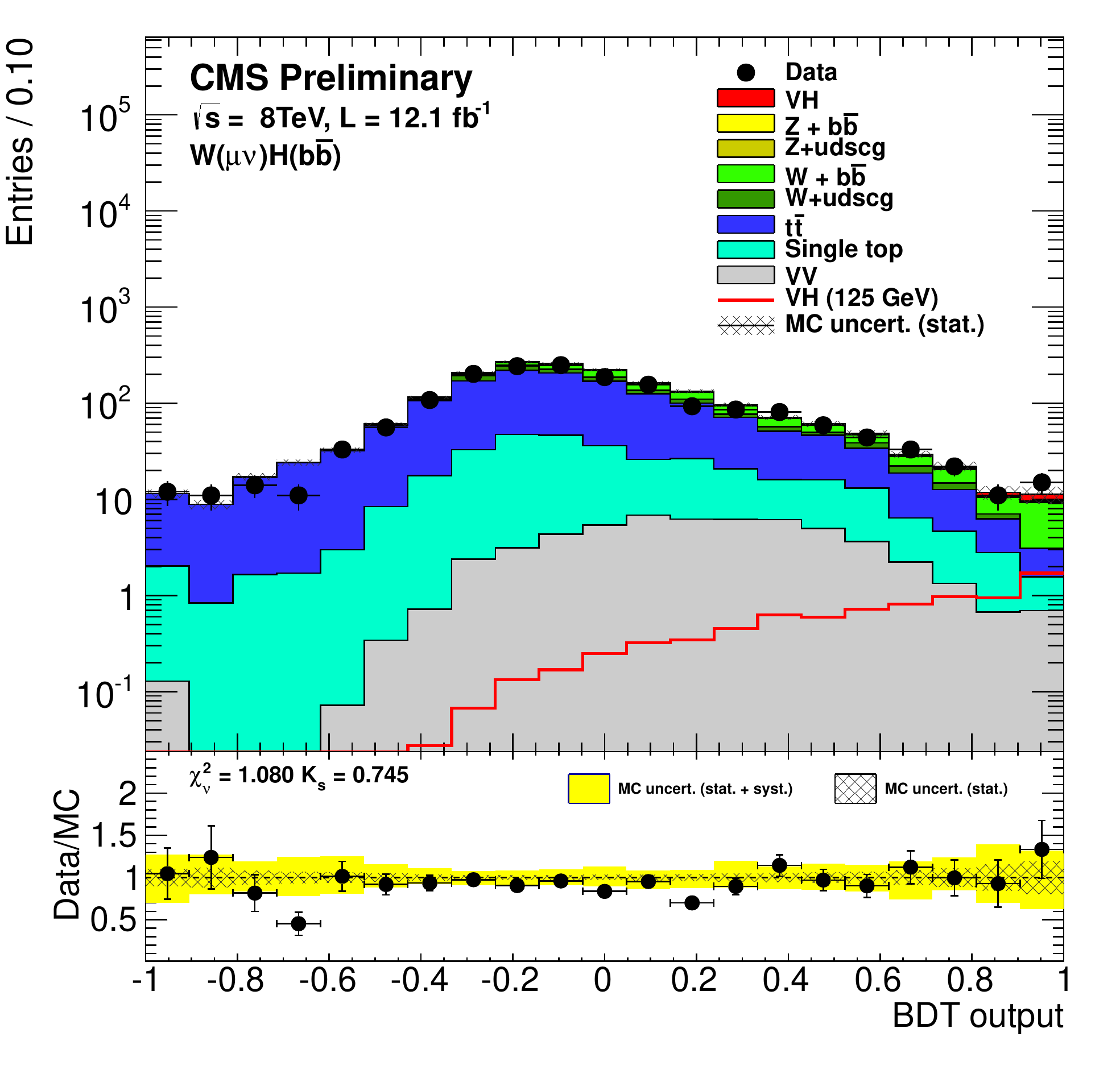}
  \includegraphics[width=0.34\textwidth,clip]{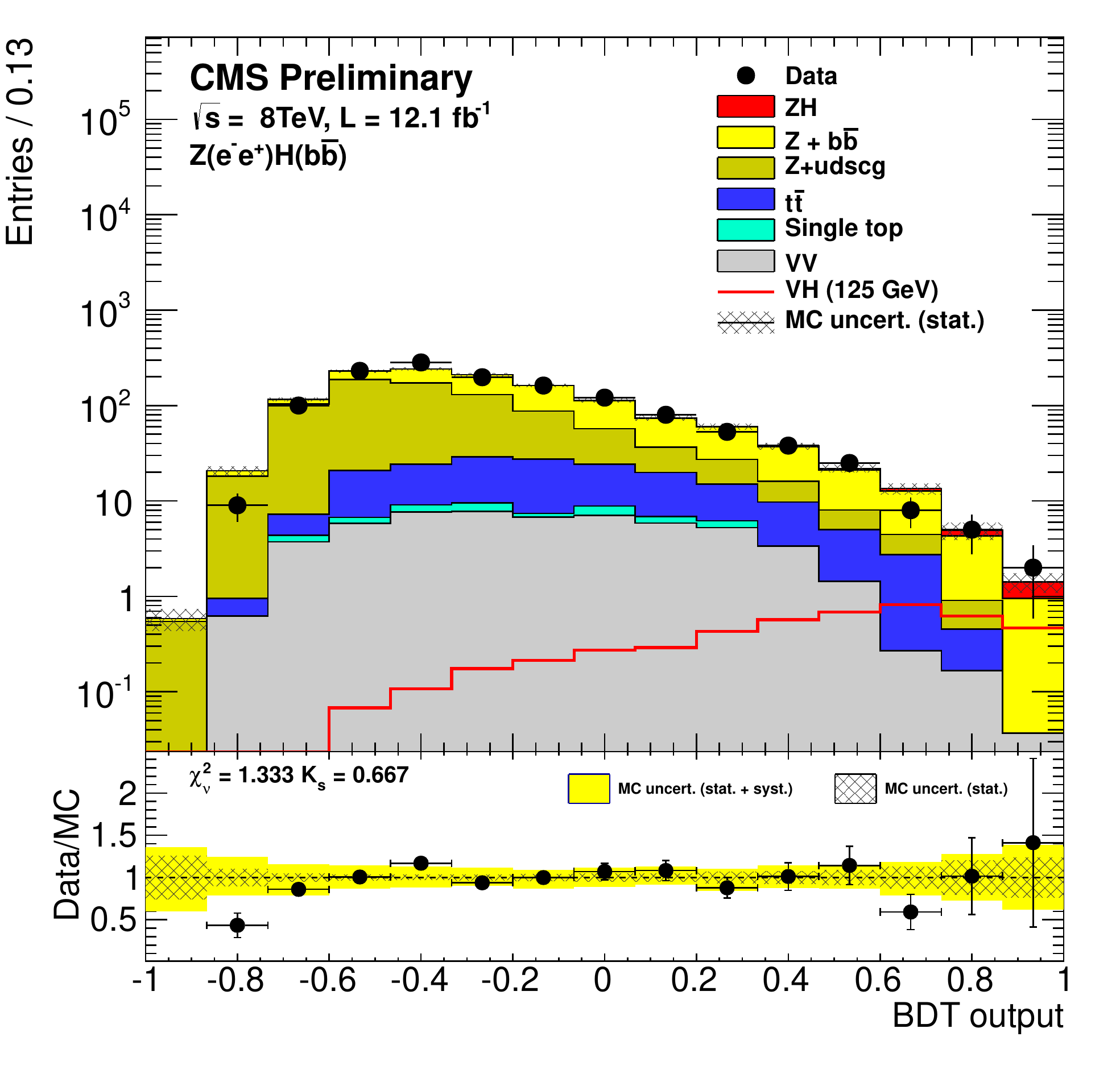}
  \caption{Examples of BDT algorithm output  distributions 
    for the CMS $VH$ analysis~\cite{CMS_VH}.
    The signal in the plots is for $m_H = $125~GeV.
    The background expectation is shown after the profile likelihood fit.
    (Left) 1-lepton channel ($W \to e \nu$), $p_{\rm T}^W > $170~GeV. 
    (Right) 2-lepton channel ($Z \to e e$), $p_{\rm T}^Z > $100~GeV. 
  }
  \end{center}
\end{figure}

For the ATLAS analysis, 
the resulting observed (expected) 95\% confidense level upper limit for a Higgs mass of
125~GeV is 1.8 (1.9) times the SM prediction,
corresponding signal strength is 
$\mu = -0.4 \pm 0.7({\rm stat.}) \pm0.8({\rm syst.}) = -0.4 \pm 1.0({\rm total})$.
For the CMS analysis, the resulting observed (expected) upper limit is 2.5 (1.2) 
times the SM prediction, which corresponds to 2.2$\sigma$ deviation 
from the no signal hypothesis and a signal strength of $\mu = 1.3^{+0.7}_{-0.6}$.

\vspace{-1.5mm}
\section{Search for the Higgs Boson in the $ttH$, $H \to b\bar{b}$}
\vspace{-1.5mm}
Searching for the signal with $ttH$ production mode is important 
to directly test top-Higgs Yukawa coupling.
SM Higgs boson production in $ttH$ process has relatively small cross section 
compared to the $VH$ process at the LHC, however, 
it should be observable with sufficient data.
For a mass of 125~GeV Higgs boson, the production cross section is:
$\sigma(ttH) = $ 0.130~pb (0.086)~pb for $\sqrt{s} = $ 8~TeV (7~TeV) at $m_H = $ 125~GeV~\cite{LHC_Xsection}.

The ATLAS analysis presently only explores the signature where the top-quark pair 
decays to one lepton+jets ($t\bar{t} \to \ell \nu q \bar{q}' b \bar{b}$, $H \to b\bar{b}$). 
The CMS analysis also considers the di-leptons final state 
($t\bar{t} \to \ell^+ \nu \ell^- \bar{\nu} b \bar{b}$, $H \to b\bar{b}$). 
The final state includes 6 jets (4 $b$-jets) for the lepton+jets channel
and 4 jets (4 $b$-jets) for the di-leptons channel.
To maximize the Higgs discovery potential, 
we categorize events based on the number of jets 
and the number of $b$-tagged jets.
The presence of a signal is tested similarly as $VH$ analyses.
ATLAS analysis fits the shape of $m_{b\bar{b}}$ invariant mass or 
scalar sum of jet $p_{\rm T}$ ($\Sigma (p_{\rm T}^{\rm jet})$), 
while CMS analysis fits the shape of the output discriminant of an 
Artificial Neural Network (ANN) algorithm which takes several kinematic input variables,
event shape and b-tagging information.
The analyzed dataset corresponds to   
an integrated luminosity of $4.7~\mathrm{fb}^{-1}$ in $pp$ collisions 
at $\sqrt{s} = 7$~TeV for ATLAS
and an integrated luminosity of $5.1~\mathrm{fb}^{-1}$ ($5.0~\mathrm{fb}^{-1}$) 
in $pp$ collisions at $\sqrt{s} = 8$~TeV (7~TeV) for CMS.

For the object selection and event selection, 
similar criteria as $VH$ analyses are used and their full details are not 
shown in this proceeding.
Full details are given in the conference notes
of the ATLAS~\cite{ATLAS_ttH} and CMS analyses~\cite{CMS_ttH}.

\subsection{Background Estimation and Analysis Optimizations}
The dominant background after event selection is $t\bar{t}$+jets production.
Due to the low signal over background ratio, it is crucial to constrain the systematic
uncertainties on the background normalization and shapes by exploiting the
various control regions.
Therefore, the normalization for the dominant $t\bar{t}$+jets contribution
is determined from final fit to the signal region and background-dominated control region.
Background-dominated regions are included for the final fit to constrain 
the background contributions in the signal regions.
After constraining the $t\bar{t}$+jets contribution, 
the theory uncertainty on $t\bar{t}$+heavy flavor is a dominant systematic uncertainty.
Both ATLAS and CMS assume approximately 50\% on the theory uncertanty 
of $t\bar{t}$+heavy flavor production.



\subsection{Results} 
In order to test the presence of the Higgs boson signal,
the test statistic is constructed to measure the compatibility of 
the data with the background only hypothesis and to derive exclusion intervals 
with the {\it CLs} method.
The ATLAS analysis uses the shape of the $m_{b\bar{b}}$ distribution 
($\Sigma (p_{\rm T}^{\rm jet})$ distribution) for $\geq$6 jets with 
$\geq$3 $b$-tags regions (for other regions),
while the CMS analysis uses the shape of output of the ANN algorithms. 
Figure~4 shows shapes of $m_{b \bar{b}}$ from the ATLAS analysis
and figure~5 shows shapes of ANN algorithm output from the CMS analysis
in the regions with the highest signal expectation. 

\begin{figure}[htbp]
  \label{fig4}
  \begin{center}
    \includegraphics[width=0.33\textwidth,clip]{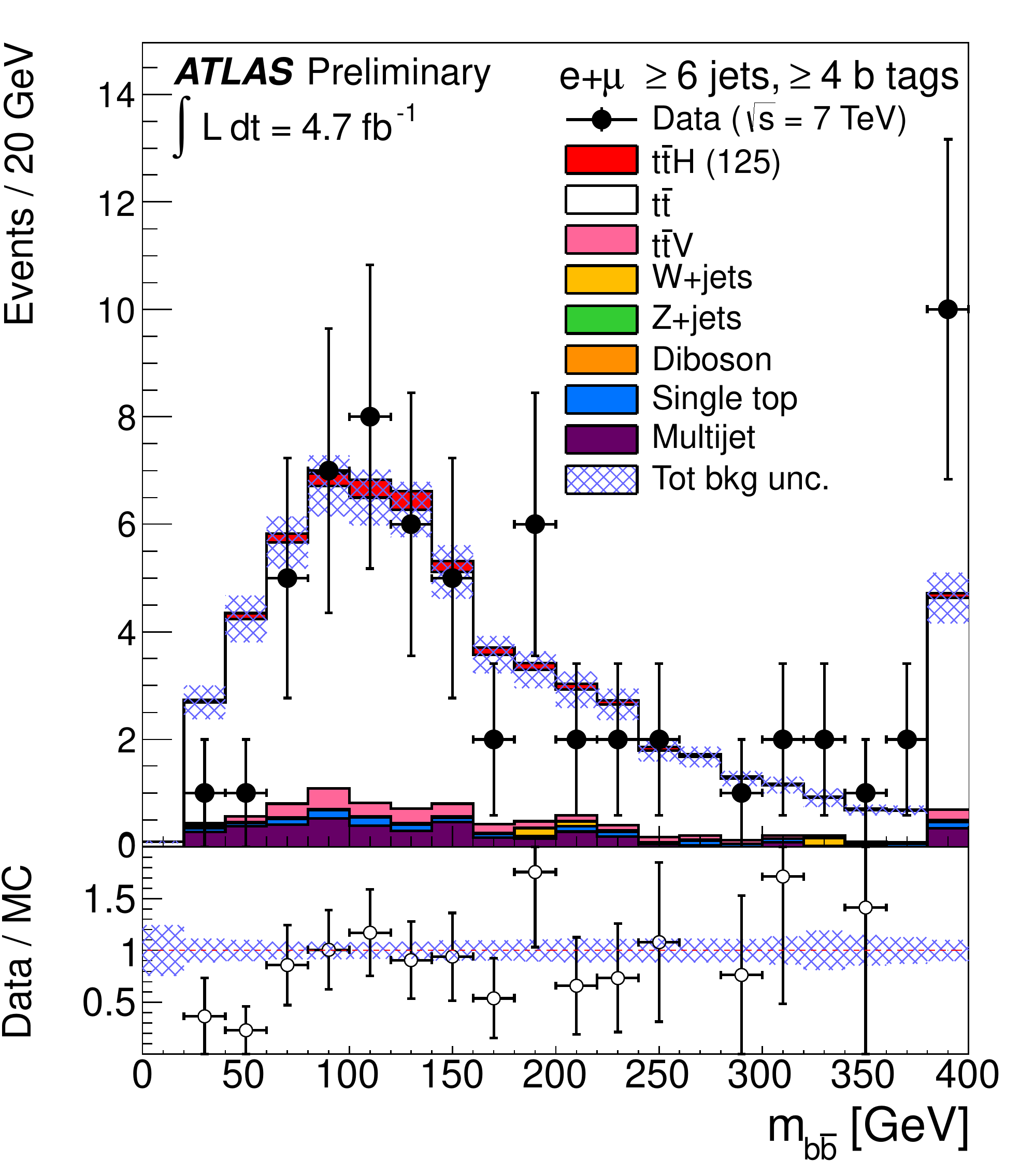}
    \includegraphics[width=0.33\textwidth,clip]{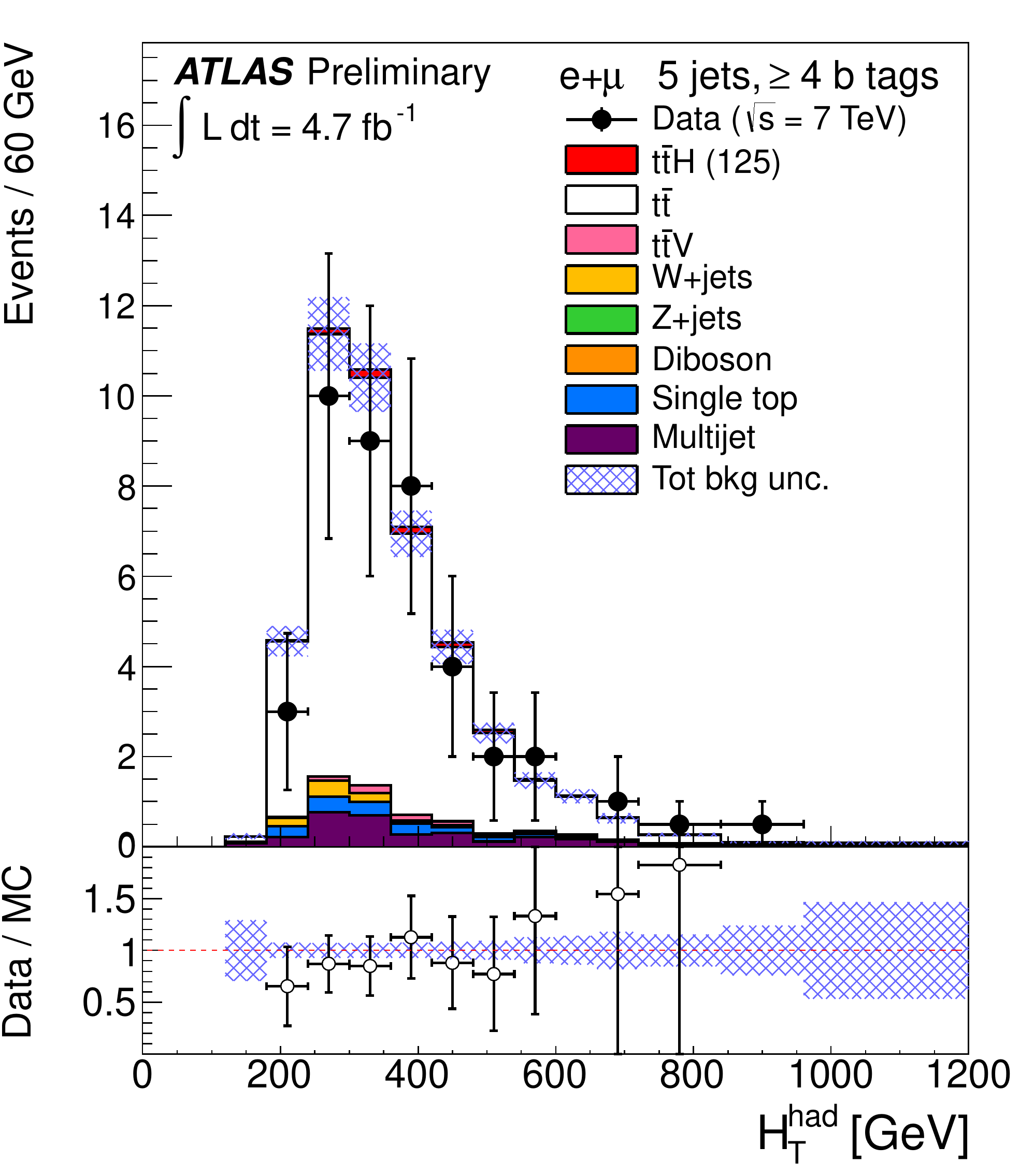}
  \caption{$m_{b\bar{b}}$ and $\Sigma (p_{\rm T}^{\rm jet})$ distribution 
    for the ATLAS $ttH$ analysis~\cite{ATLAS_ttH}.
    (Left) $\geq$6 jets ($\geq$4 $b$-tags) category. 
    (Right) 5 jets ($\geq$4 $b$-tags) category. 
    The signal in the plots is for $m_H = $125~GeV.
    The background expectation is shown after the profile likelihood fit.
  }
  \end{center}
\end{figure}

\begin{figure}[htbp]
  \label{fig5}
  \begin{center}

  \includegraphics[width=0.3\textwidth,clip]{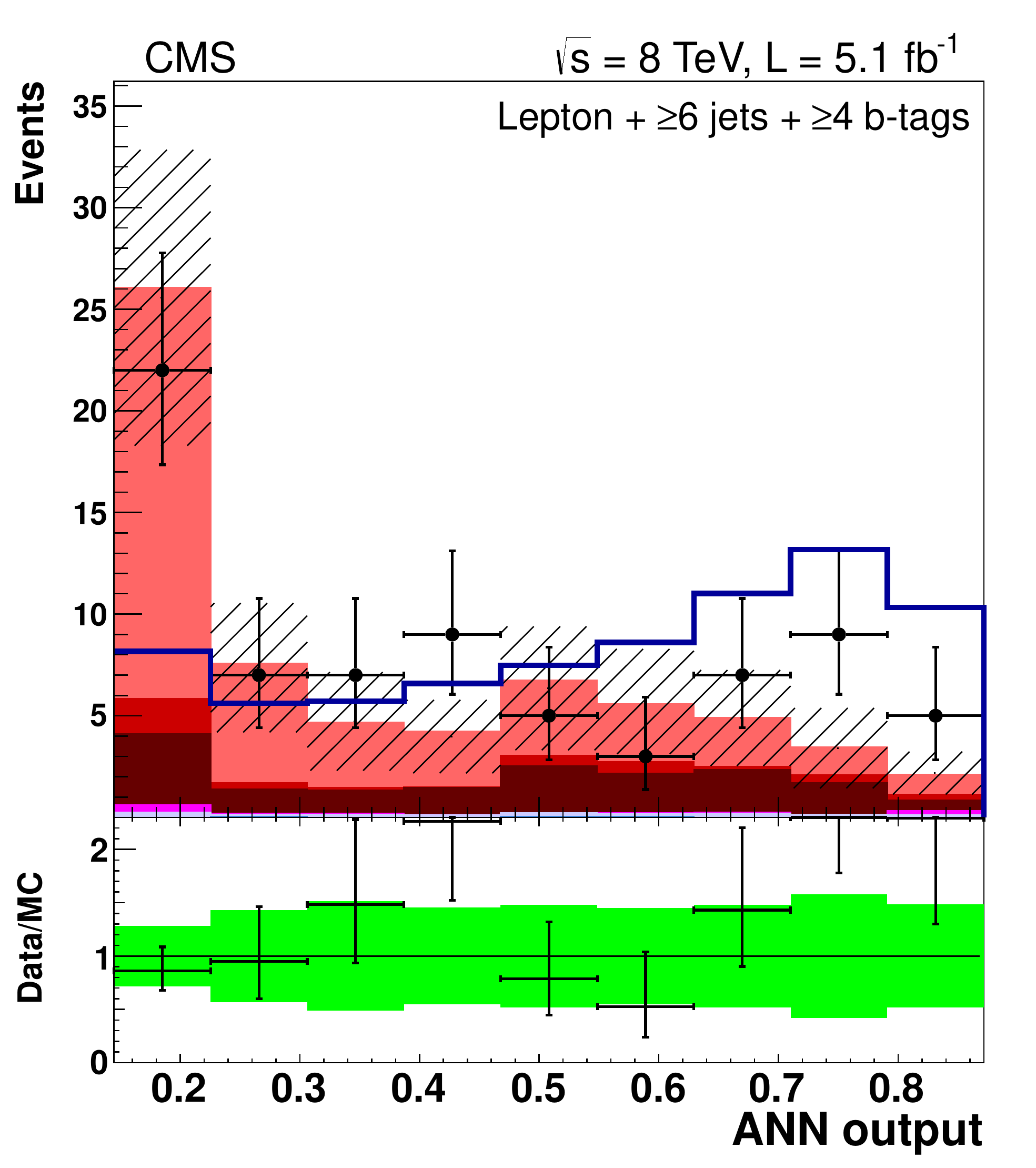}
  \includegraphics[width=0.3\textwidth,clip]{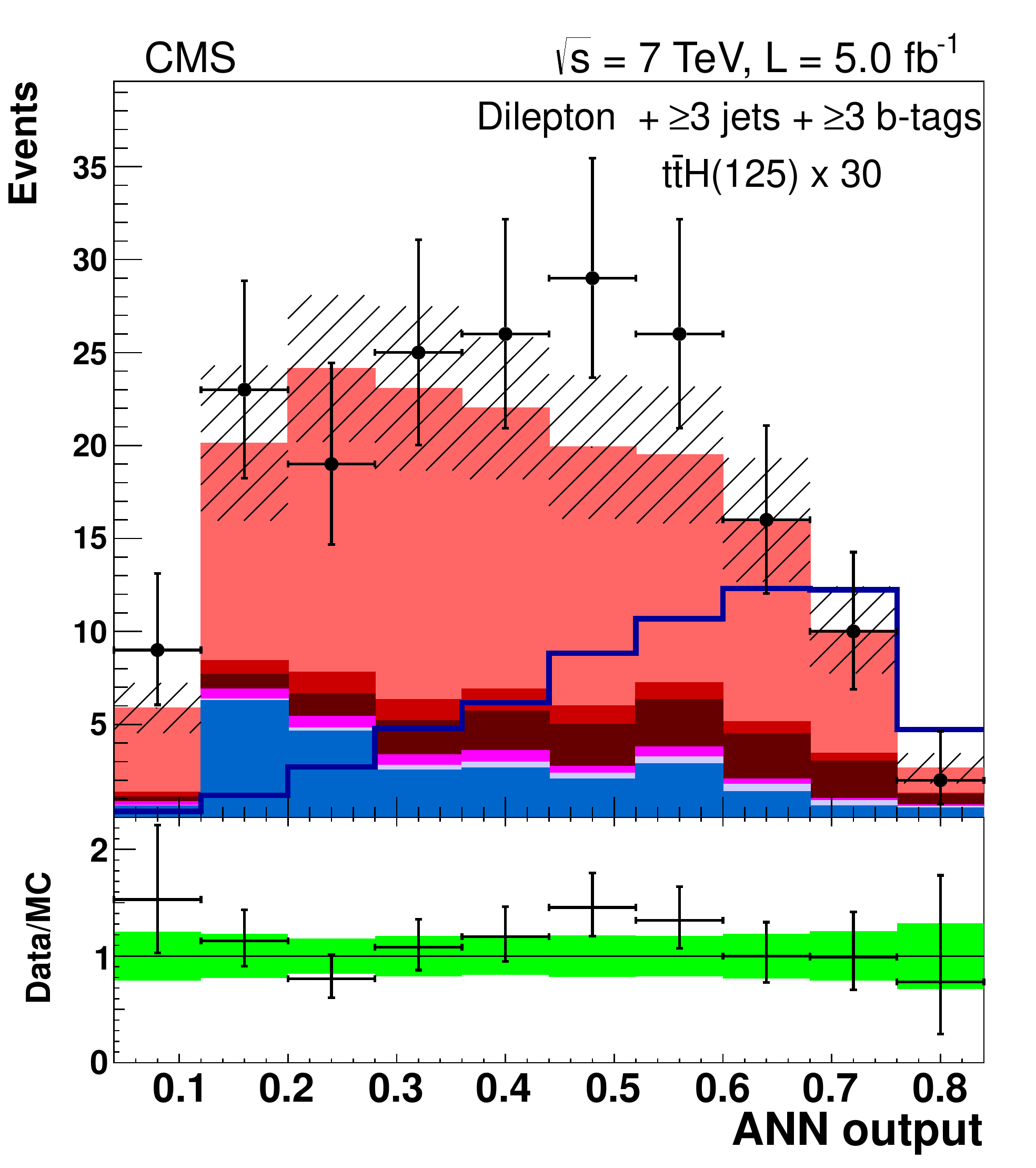}
  \includegraphics[width=0.2\textwidth,clip]{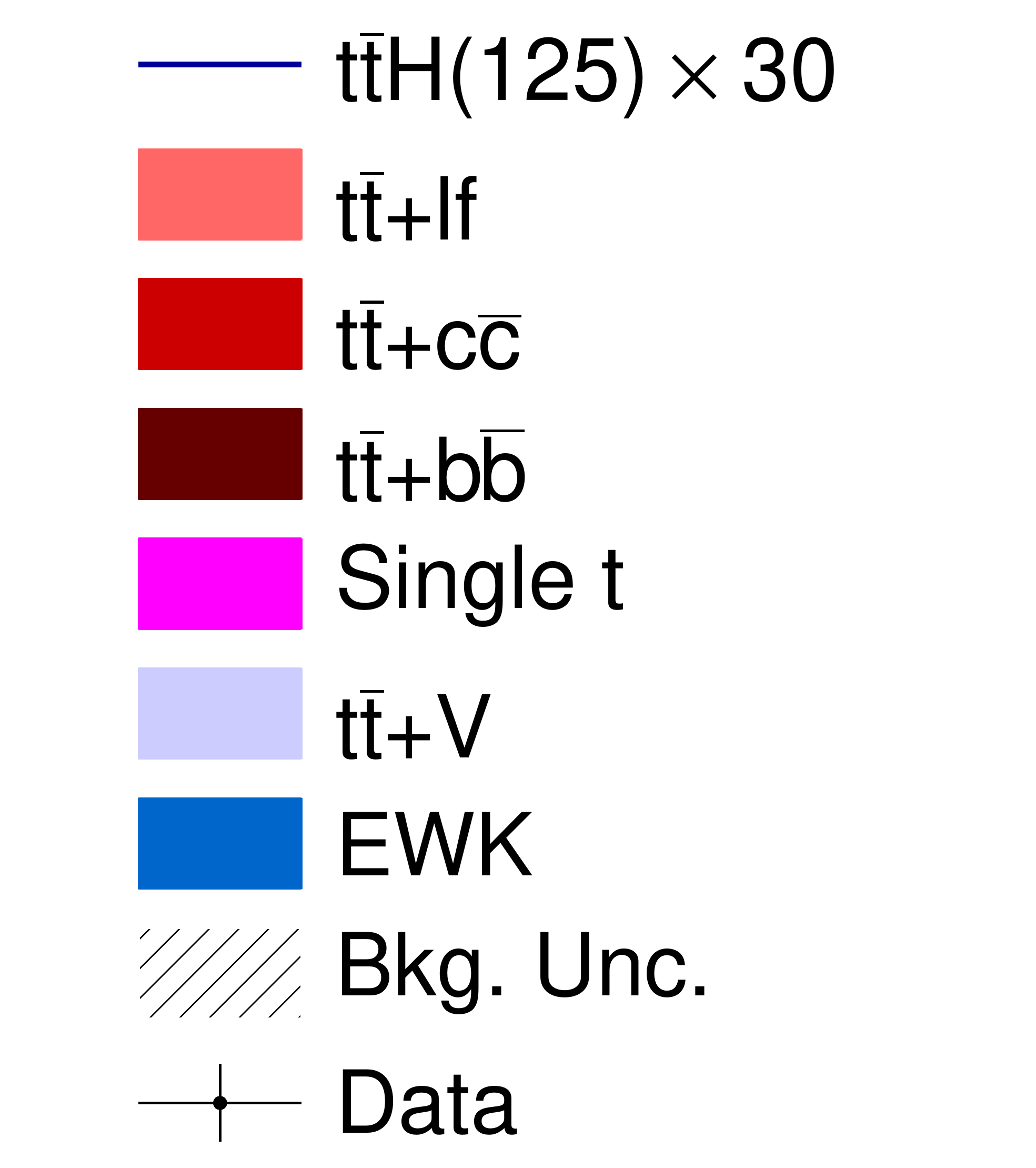}
  \caption{ANN algorithm output distributions 
    for the CMS $ttH$ analysis~\cite{CMS_ttH}.
    (Left) lepton+jets channel, $\geq$6 jets ($\geq$4 $b$-tags) category.
    (Right) di-leptons channel, $\geq$3 jets ($\geq$3 $b$-tags) category.
    The signal in the plots is for $m_H = $125~GeV.
    The background expectation is shown after the profile likelihood fit.
  }
  \end{center}
\end{figure}

In the CMS $ttH$ analysis, all Higgs boson decay modes are taken into account, 
although the analysis is optimized for the decay mode $H \to b\bar{b}$.
The limits are expressed as the multiple of the SM Higgs boson production
cross section which is excluded at 95\% confidense level 
for each value of the Higgs boson mass.
For the ATLAS analysis, 
the resulting observed (expected) upper limit for a Higgs boson mass of
125~GeV is 13.1 (10.5) times the SM prediction.
For the CMS analysis, 
the resulting observed (expected) upper limit for a Higgs boson mass of
125~GeV is 5.8 (5.2) times the SM prediction.


\vspace{-1.5mm}
\section{Summary and Prospects}
\vspace{-1.5mm}
In this proceeding searches for the SM Higgs boson using the $H \to b\bar{b}$ decay mode by both 
the ATLAS and CMS experiments have been presented.
To validate the analysis method, a search is performed for the well-known SM diboson signal 
($VZ \to V b\bar{b}$) and a clear excess is observed which is in consistent with the SM prediction. 
The searches for the $VH$ signal process performed by ATLAS and CMS result in sensitivities 
which are close to the SM prediction. 
While the ATLAS experiments sets an upper 95\% confidence limit at 1.8 times the Standard Model signal expectation, 
the CMS experiment sees a first indication of a signal, which corresponds to a 2.2$\sigma$ excess 
at $m_H = $ 125~GeV with respect to the background only expectation.
$H \to b\bar{b}$ decays in the $ttH$ production mode are also searched for but 
neither of the experiments observes an excess. 
Therefore upper limits on the SM Higgs boson production cross section times the branching ratio are set.

Both ATLAS and CMS have yet to analyze the full 2012 dataset. 
In addition, further work is ongoing in both collaborations to achieve improvements well beyond the increase 
in statistical sensitivity due to the addition of new data. 
The new analyses including the full 2012 dataset might therefore lead to an observation of the $H \to b\bar{b}$ mode, 
if this turns out to be realized in nature.

\end{document}